# Geo-political conflicts, economic sanctions and international knowledge flows

Teemu Makkonen[#] and Timo Mitze[§,*]

[#] Faculty of Social Sciences and Business Studies, Karelian Institute, University of Eastern Finland (UEF), Yliopistokatu 2, Aurora B, Joensuu, Finland

[§] Department of Economics, University of Southern Denmark (SDU), Campusvej 55, DK-5230 Odense, Denmark

* To whom correspondence should be addressed. E-mail: tmitze@sam.sdu.dk

**Abstract:** We address the question how sensitive international knowledge flows respond to geo-political conflicts taking the politico-economic tensions between EU-Russia since the Ukraine crisis 2014 as case study. We base our econometric analysis on comprehensive data covering more than 500 million scientific publications and 8 million international co-publications between 1995 and 2018. Our findings indicate that the imposition of EU sanctions and Russian counter-sanctions from 2014 onwards has significant negative effects on bilateral international scientific co-publication rates between EU countries and Russia. Depending on the chosen control group and sectors considered, effect size ranges from 15% to 70%. Effects are also observed to grow over time.

**Keywords:** knowledge flows; scientific co-publications; geo-political conflicts; economic sanctions; Ukraine crisis; EU-Russia

JEL classification: C23, F51

## 1. Introduction

This paper investigates the impact that geo-political conflicts, particularly economic sanctions, have on knowledge flows between conflict parties. While the use of sanctions as foreign policy tools has grown over time, our scientific understanding of their intended and unintended socio-economic effects is still in the making (e.g., Hovi et al., 2005, Hufbauer & Jung, 2020; Felbermayr et al., 2020, 2021). Different from previous work, which has predominately studied implications for bilateral trade, FDI and labor mobility following a sanctions period, our focus is set on the development of international knowledge exchange measured through scientific co-publications. We argue that a focus on the latter is well deserved given its importance to spur innovation and economic development in general (Peri, 2005; Coscia & Wang, 2016; Chen et al., 2019; Coscia et al., 2020) and in addressing new global challenges, such as climate change and the COVID-19 pandemic.

We use the Ukraine crisis as empirical case here and conduct a comprehensive quantitative assessment on its impact for knowledge flows between EU countries and Russia as major conflict parties. The Ukraine crisis, which is associated to be one of the largest geo-political threats in the recent past (Caldara & Iacoviello, 2018), was triggered by a chain of political events in the Ukraine which culminated in the annexation of the Crimea by the Russian Federation. Seeing this annexation as a violation against international law, the European Union (EU) – coordinated with other Western countries such as the USA – has imposed several restrictive measures against the Russian Federation as of March 2014 including travel bans, asset freezes as well as sanctions on international trade for dual use goods and financial flows (European Council, 2021). The Russian Federation has met these sanctions with reciprocal counter-sanctions including an import embargo on selected goods (mainly food items) produced in the EU.

While no explicit sanctions on scientific knowledge exchange have been imposed by the involved conflict parties, there are several reasons to expect a negative impact of sanctions on knowledge flows. First, funding access to EU framework programmes has been restricted for Russian researchers; second, researcher mobility between the EU countries and Russia has been restricted and, third, projects involving technologies with potential military applications have been suspended. For empirical effect identification, we follow the growing number of empirical contributions that exploits the quasi-experimental nature of imposed sanctions following a sudden political crisis to identify their likely causal effects (e.g., Ahn & Ludema, 2020, Bělín and Hanousek, 2020, Crozet et al., 2021). Exploiting this exogenous variation in a rigorous treatment effect analysis shall help us to answer to the specific research question of this paper: did sanctions and counter-sanctions between the EU and Russia lead to a significant decline in EU-Russia scientific collaboration?

The revealed costs of escalating sanctions between the EU and Russia are mutual with detrimental effects being previously observed for bilateral trade, cross-border capital flows and the movement of people between the EU and Russia. As such, the EU-Russian trade relationship has been observed to experience a general decline in bilateral export and import flows as consequences of the sanctions, resulting in trade losses for both parties (Giumelli, 2017; Crozet & Hinz, 2020; Doornich & Raspotnik, 2020; Crozet et al., 2021). For Russia, the negative impacts of sanctions are, naturally, greater in sectors that are dependent upon Western inputs (Ahn & Ludema, 2020). Other evidence on asymmetric effects points to a higher decline in trade flows associated with the Russian counter-sanctions on EU food products compared to the trade restrictions imposed by the EU (Bělín & Hanousek, 2020).

While this appears to be compelling evidence for a negative sanctions effect of international trade, one potential confounding factors is that the decline in trade coincided with a weakening of the Russian ruble and falling oil prices potentially overlapping with the sanctions effect (Dreger et al., 2016). Beyond the trade dimension, instabilities caused by the sanctions are also found to have negative effects on the investment climate between the EU and Russia resulting in a significant drop in the inflow of foreign direct investments to Russia (Liuhto et al., 2017) and a reduction in the movement of people as evident from tourism flows between the EU and Russia (Ivanov et al., 2017; Makkonen et al., 2018; Prokkola, 2019).

Building on these earlier findings, we adopt a quasi-experimental research design to assess if scientific knowledge flows have been similarly affected by the crisis as the international flows of goods, capital, and people and – if so – seek to identify, which scientific fields have been the most affected. Specifically, we apply flexible difference-in-difference (DiD) estimation for quantifying average and dynamic treatment effects over time. Regressions are carried out within a gravity model framework for bilateral knowledge flows between countries and are based on aggregate and field-specific data on international scientific co-publications between the EU, Russia, and other BRICS countries. This framework allows us to control for confounding factors in the relationship between sanctions and knowledge flows by including time-varying covariates and a multidimensional fixed effects structure to arrive at robust treatment effect estimates (Balazsi et al., 2018).

While bilateral scientific co-publications between EU countries and Russia are considered as knowledge flows subject to treatment (i.e. economic sanctions), intra-EU, EU-BRICS (minus Russia) and Russia-BRICS bilateral knowledge flows are used to establish counterfactuals, i.e. predict trajectories of EU-Russia scientific collaborations if sanctions would have not occurred. The selection of our control group in terms of using other BRICS (Brazil, Russia, India, China, South Africa) countries to establish counterfactuals, is mainly driven by two motives: First, countries should be similar to Russia in terms of their economic development, their underlying science system and their pre-sanctions international scientific relations to the EU, and they should not have imposed sanctions on Russia during the Ukraine crises. Considering these selection principles, we argue that the rapid imposition of EU sanctions and Russian counter-sanctions, which quickly entered into force after the Russian annexation of Crimea in March 2014, serves as an exogenous source of variation in the data. Following Bělín and Hanousek's (2020) rationale, we thus make the point here that the reciprocal sanctioning progression following the Ukraine has a solid claim to be quasi-experimental due to the geo-political considerations that drove the imposition of sanctions and their timing. Importantly, we also argue that scientific collaborations are less sensitive to short-run economic developments, such as in the case of Russia a weakening of the Russian ruble and falling oil prices, which may render our analysis more robust than prior investigations on international trade and investment flows.

## 2. Drivers and impediments of knowledge flows

The potential impacts of the sanctions on scientific collaboration between the EU and Russia have remained largely unexplored. What is known based on earlier literature is that EU and Russian officials have traditionally been supportive of the idea of close collaboration between scientists from the EU and Russia (Yegorov, 2009). However, while the sanctions have not explicitly excluded Russia from participating in international research efforts (Schiermeier, 2018), the contemporary geo-political climate seems to have impaired scientific cooperation between the EU and Russia. This is due to: i) ***decreased mobility*** between the EU countries and Russia and reduced opportunities for Russian academics to work in international laboratories, ii) ***decreased collaborative research funding*** due to restricted access of Russia to EU framework programmes and iii) the suspension of a number of international research projects due to a significant proportion of electronic components for R&D and modern equipment and technology falling under the EU export sanctions as they are considered as ***"dual use technologies"*** with potential military applications. Consequently, the sanctions are leading Russian researchers to find partners from non-Western counterparts such as the BRICS countries (Kotsemir et al., 2015), which have not imposed any sanctions against but rather shown support for Russia. We give a short account of these three key factors in the following.

***Decreased mobility:*** As stated by Fernández-Zubieta and Guy (2010) increased international researcher mobility and the subsequent greater interaction of research-related personnel are key factors in facilitating knowledge and technology transfer. That is, mobility of researchers is seen as one of the main channels of scientific collaborations and knowledge flows (Atta-Owusu, 2019), since international researcher mobility is expected to facilitate the exchange of both scientific as well as institutional and cultural knowledge (Petersen, 2018). The process is self-reinforcing as an initial mobility period often launches a cumulative process of subsequent researcher mobility and increased collaboration (Jöns, 2009). Trippl (2013) has formally analysed this proposed connection between researcher mobility and knowledge flows. With her "knowledge links model", she evidenced that the mobility of academic scientists contributes to knowledge transfer and even further to the growth of the regions involved. The process is propelled by increased further mobility (the mobile researcher's former team members or students may follow her or him) and via the maintaining of the ties with partners in the country of origin of the researcher leading to reciprocal forward and backward knowledge flows (see also Schiller & Diez, 2012; Atta-Owusu, 2019).

Several other empirical studies (e.g. Gibson & McKenzie, 2014; Appelt et al., 2017; Aman, 2018; Aykac, 2021) have since then corroborated the view that international researcher mobility increases knowledge flows in both directions between sender and receiving countries. Thus, a drop in academic mobility can be expected to reflect negatively on knowledge flows between Russia and the EU. While academic mobility is not as such targeted via the economic sanctions there seems to be a significant overlap in the slowdown of EU-Russia academic mobility and the sanctions year of 2014. For example, according to the figures of Education Statistics Finland (2021) academic mobility from Russia to Finnish institutes of higher education plummeted from the ca. 2000 mobility periods in 2013 to only 1000 mobility periods in 2015 (see Figure A1 in the supplementary appendix for details).

***Decreased collaborative research funding:*** The impact of funding on research collaboration is commonly perceived as positive (Ubfal & Maffioli, 2011). Therefore, participation in, for example, EU funded projects has been remarked to create multinational networks that will (potentially) give rise to international knowledge flows (Di Cagno et al, 2016). In fact, by using scientific collaboration in EU funded research projects under the FP programmes over the period 1994–2005, Di Cagno et al. (2014) have found that participation in EU funded projects is indeed an important channel of knowledge flows. More precisely, the creation of collaborations among the different countries was shown to be affected by their participation in joint research projects. These collaborations, in turn, acted as the channels through which knowledge flows between sender and receiving countries. As such, participation in multinational joint research projects, is one of the mechanisms that facilitates knowledge to flow between countries (see also Caloghirou et al., 2006).

The most recent figures on Russian participation in Horizon 2020 so far has significantly declined in comparison to FP7, which is partly explained by the new funding rules of Horizon 2020 (whereby Russian participants are not automatically eligible for EU funding). However, according to the European Commission (2018) the drop is also due to the current sanctions against Russia and the concerns about cooperation in some technologically sensitive areas (in particular, the dual use technologies) that have led some scientists to think that Russian partners in Horizon 2020 collaborative projects would not be welcome. It thus seems that the sanctions have hampered the flow of knowledge between the EU and Russia through decreased collaborative funding and joint research projects.

***"Dual use technologies":*** While there are very few studies investigating knowledge flows in dual use technology sectors, the rationale behind knowledge sharing between military industries and the civilian sectors lie in the possibility of creating synergies for innovation and social benefits (Acosta et al., 2020). However, for obvious reasons such dual use technologies having potential military applications have been directly targeted by the EU-Russia sanctions: the trade of such goods is restricted (European Council, 2021) and the cooperation with the Russian military-industrial complex and their suppliers is banned (Volovik, 2020). The restrictions to research cooperation with foreign partners naturally thwart the mundane research practices of knowledge sharing in subject areas linked to dual use technologies (Krige, 2014), which can be expected to result in decreased knowledge flows between Russian and the EU.

## 3. Empirical Strategy

We use the above arguments as theoretical priors in our empirical investigation of the impact of the start of the Ukraine crisis and the imposition economic sanctions on scientific collaboration. We do so by running treatment effect regressions within the framework of a gravity model. Gravity models are regarded as a workhorse approach for the analysis of spatial interaction data to identify treatment effects associated with factors fostering international economic exchange, e.g., through trade integration agreements, or in reducing international relations, e.g., through war, conflicts, and sanctions (Baier & Bergstrand, 2007; Dai et al., 2014). Gravity models also have a long tradition in the empirical analysis of factor mobility (capital and migration flows) and have recently also been adopted for the analysis and modelling of knowledge flows and innovation cooperation (Peri, 2005; Picci, 2010; Keller & Yeaple, 2013). In its triple index form, the basic gravity equation used in our empirical investigation can be written as

$$(1) \qquad ICPR_{ijt} = \beta_0 \cdot \mathbf{X}_{\mathrm{ijt}}^{\beta_1'} \cdot DIST_{ij}^{\delta} \cdot SANC_{ijt}^{\gamma} \cdot e^{\theta_i d_i + \theta_j d_j + \lambda_t z_t} \cdot \varepsilon_{ijt},$$

where $ICPR$ is the international co-publication rate for country pair $i\,j$ at time $t$ as our outcome variable of interest is regressed against a vector of time-varying controls characterising the attractiveness of a country pair for scientific collaboration ($\mathbf{X}_{\mathrm{ijt}}$). $DIST$ is a measure for geographical distance working as an impediment to international research interaction in the logic of the gravity model and $SANC$ is a measure that captures changes in foreign policy for a specific treatment group over time (i.e. in our case the period of economic sanctions directly affecting EU-Russia country pairs); $d_i$ and $d_j$ are a set of dummies for countries $i$ and $j$ (i.e. a set of country-fixed effects) that serve as a proxy for multilateral resistance terms (Fally, 2015) and capture latent country-specific characteristics such as political institutions, the national innovation system etc. Further, $z_t$ are dummies for individual sample years to cover time trends and shocks (e.g. business cycle movements) common to all country pairs and $\varepsilon_{ijt}$ is a well-behaved identical and independently distributed *i.i.d.* error term; finally, $\beta_0$, $\beta_1'$, $\delta$, $\gamma$ , $\theta_i$, $\theta_j$ and $\lambda_t$ are parameter (vectors) to be estimated.

In line with the gravity mode literature, we follow Rose's approach (Rose, 2004) and construct time-varying control variables ($\mathbf{X}_{\mathrm{ijt}}$) in such a way that the represent the product of values for countries *i* and $j$; for instance, $GDP_{ij}$ enters the gravity equation as $GDP_{ijt} = (GDP_{it} \cdot GDP_{jt})$, treating variables as pairwise (dyadic) measures. Similarly, we also include the product of publications ($Pub$) from country *i* and $j$ at time *t* in the set of controls. As an extension to the basic gravity equation shown in equation (1), we also estimate the gravity equation including a more extensive set of country-pair dummies $\theta_{ij}d_{ij}$, which additionally capture structural differences between individual pairs of countries such as bilateral distance, common language etc. Finally, we also interact the included time trends with the country dummies, i.e., $\Psi_{it}(\ d_i \cdot z_t), \Psi_{jt}\left(\ d_j \cdot z_t\right)$ for which the coefficients $\Psi_{it}$ and $\Psi_{jt}$ capture time-varying country heterogeneities in co-publication activities potentially confounding with the treatment effects of sanctions. We refer to this complex set of dummies as multidimensional fixed effects (FE) structure.

Our treatment indicator is $SANC$, which captures the spatio-temporal variations in sanctions. variable is specified as a binary dummy that takes a value of 1 for all country pairs that involve Russia and any EU country from 2014 onwards; it is zero otherwise. This means that scientific collaboration within the EU, between the EU and the remainder of BRICS countries (excluding Russia) as well as within the group of BRICS countries (including Russia) serve as potential comparison settings for the temporal evolution of EU-Russia scientific collaboration in the logic of difference-in-difference (DiD) estimation. In the case that the EU-Russian scientific collaborations fall short of those in the control groups, then the regression coefficient for $SANC$ takes a value of $\gamma$<0 and is taken as evidence for a negative effect of economic sanctions on knowledge flows. To test for the presence of early anticipation and gradual phasing-in treatment effects, i.e., dynamic treatment effects, we also estimate gravity model specifications with separate treatment indicators for individual sample years $\sum_{n=-N}^{N} SANC_{n,ijt}^{\gamma_n}$, where [-*N*,*N*] describes the maximum number of leads and lags relative to the imposition of the sanctions in 2014 included in the gravity model; each parameter $\gamma_n$ measures the effect for a given lag/lead year.

Our most flexible DiD regression specification with multidimensional fixed effects (FE) structure thus takes the form[1]

(2) $$ICPR_{ijt} = \beta_0 \cdot \mathrm{X}_{\mathrm{ijt}}^{\beta_1'} \cdot DIST_{ij}^{\delta} \cdot \sum_{n=-N}^{N} SANC_{n,ijt}^{\gamma_n} \cdot e^{\theta_{ij} d_{ij} + \Psi_{it}(d_i \cdot z_t) + \Psi_{jt}(d_j \cdot z_t)} \cdot \varepsilon_{ijt}.$$

In terms of estimation, earlier empirical work on the gravity model (of trade) has predominately applied Ordinary Least Squares (OLS) to log-transformed variables. However, recent estimator comparisons have shown that the Poisson regression with over-dispersion belonging to the family of Generalised Linear Models (GLM) may be a better choice (McCullagh & Nelder, 1989; Wooldridge, 2010). While earlier studies have proposed a Poisson Pseudo Maximum Likelihood (PPML) estimator for gravity models in their multiplicative form (Santos Silva & Tenreyro, 2006), recent evidence has shown that the PPML is also a proper estimator choice in DiD-type settings (Ciani & Fisher, 2019; Besedeš et al., 2021). Therefore, we estimate the gravity model equation shown in equation (1) both by OLS for the pooled (POLS) and fixed effects (FE) linear panel model and by PPML. The main difference between the POLS and FE model is the multidimensionality of the included fixed effects. Since the model is estimated based on dyadic data for *ij* country pairs, we only consider regression specifications including country-pair fixed effects as a true FE model (specifications controlling for country-level fixed effects are classified as pooled specification with partial fixed effects). In the case of the linear OLS estimations, we choose a semi-log specification applying a logarithmic transformation to right-hand-side variables in equation (1) but do not transform the outcome variable defined as a percentage rate.

## 4. Data and Stylized Facts

***Data and Variables.*** Scientific co-publications are chosen here as key indicator for scientific knowledge flows, as done in a number of other recent papers on international research collaboration (e.g. Newman, 2004; Gui et al., 2018; Zhang et al., 2018; Wagner et al., 2019), for two reasons: 1) Scientific publications are the dominant form of diffusing scientific knowledge among the academia (Nelson, 2009) and 2) bibliometric analysis of scientific co-publications has been for long one the most common ways of measuring (international) scientific collaboration (Melin & Persson, 1996). We consider a scientific co-publication to be international if it is authored by at least two authors located in two different countries. An article having authors from several countries will show as a co-authored publication for each country-pair (Makkonen & Mitze, 2016). For example, an article having authors from Russia, Finland and Germany will show in our data as an international scientific co-publication between 1) Russia and Finland, 2) Russia and Germany as well as 3) Finland and Germany. All data on scientific co-publications were gathered on December 2019 from the Web of Science (WoS) Core Collection maintained by Clarivate Analytics and include the Science Citation Index Expanded (SCI-EXPANDED), the Social Science Citation Index (SSCI), the Art & Humanities Citation Index (A&HCI) and the Emerging Sources Citation Index (ESCI). The search procedure utilised the address field and simple Boolean logic – for example, “ADDRESS: (Russia) AND ADDRESS: (Austria)”.

[1] See also, Callaway and Sant'Anna (2020) for a general discussion of DiD estimation with multiple time periods.

The data collected on international scientific collaborations cover all EU-28 (including the UK) and BRICS countries. The latter group includes Russia together with Brazil, India, China and South Africa. BRICS countries (minus Russia) serve as essential control units in the DiD estimations. The underlying data structure thus consists of 33×32=1,056 country pairs for 1995–2018 (=25,344 country pair × year observations). As outlined above, we have decided to use other BRICS countries than Russia as essential control units as they are similar in terms of their economic development level and, importantly, can be treated as statistically independent from the treatment, i.e., none of these countries have introduced economic sanctions as strategic policy response to the Ukraine crisis. We also investigated if other policy actions may have influenced international scientific collaborations between the EU and BRICS countries during the treatment period but did not find evidence for any intervening event that may bias our estimation results from an ill-defined control group. While we also considered other countries for the inclusion in our county sample, such as the USA, Canada, Australia or Turkey, but all these countries have equally introduced sanctions on Russia during the Ukraine crisis.

We are aware of the caveats related to the use of bibliometric data. These include 1) the sensitivity of search queries to spelling errors, inconsistencies in country names in the address field of articles and double affiliations (i.e. one author having two or more addresses in different countries). Moreover, 2) databases like WoS tend to favour English language journals (Aldieri et al., 2018). We have carefully accounted for potential spelling errors and most obvious sources of inconsistencies; for instance, Northern Ireland and New South Wales, were considered in the search procedures (utilising Boolean logic) for Ireland and the United Kingdom. With regard to the issue of double affiliations, we argue that an author working in two (or more) countries can be expected to establish international cooperative links and increase knowledge flows between the two (or more) organisations she is affiliated to (Trippl, 2013). And finally, we argue that the contribution of non-English articles to the analysis of international knowledge flows remains limited (Lovakov & Agadullina, 2019) and is not expected to systematically bias our empirical results.

In the absence of direct restrictions to international scientific collaboration it is *a priori* unclear what types of knowledge flows are likely affected by the sanctions. To identify all possible (side) effects of economic sanctions on international knowledge flows, data gathering has been done at different levels of aggregation. Firstly, we have collected information on the total number of co-publications across all subject categories. For the sample period 1995–2018, this corresponds to a total of 506.83 million publications and 8.32 million international co-publications. Since, this broad definition of knowledge flows may create noise in the statistical analysis, secondly, we also collected publications for pre-defined WoS research fields of i) Technology (128 million publications, 1.39 million international co-publications), ii) Physical Sciences (196 million publications, 4.21 million international co-publications), iii) Life Sciences and Biomedicine (208 million publications, 3.30 million international co-publications, and iv) Social Sciences, Arts and Humanities (331 million publications, 3.16 million international co-publications).

The research fields 1) Social Sciences and 2) Humanities and Arts have been merged given that the number of international co-publications in the latter subfield was comparably smaller than in the other WoS research fields. Thirdly, we have compiled a fine-tuned list of 46 WoS subject categories closely related to the EU's export restrictions for dual-use technologies (152 million publications, 2.21 million international co-publications). The matching between the dual-use technologies and WoS subject categories was made based on Scalia et al.'s (2017) description of dual-use technologies. The matching was done by the authors (see Table A1 in the supplementary appendix for details). All field-specific data are utilised as a supplementary figure to the analysis of aggregate data with the goal to provide robust results.

Our key outcome variable is defined as the international co-publication rate for a given country pair *ij* at time *t*, which is expressed as the number of international co-publications $(Copub_{ijt})$ measured as percentage share of the average number of total publications $(Pub)$ in countries *i* and *j* at time *t* as

$$ICPR_{ijt} = \left(\frac{Copub_{ijt}}{1/2\, Pub_{it} + 1/2\, Pub_{it}}\right) \cdot 100.$$

As shown in Table 1, the sample mean of the international co-publication rate is 1.27% but it is observed to vary between 0 and 18% across country pairs in the overall sample. Significant differences in the shares of international co-publications can be observed across scientific fields, with the highest average international co-publication rate being observed for Physical Sciences (1.92%) followed by the dual use category (1.33%). In the latter category, the maximum value for the international co-publication rate can be up to 59% underlining the relative importance of international knowledge exchange and diffusion for scientific progress in this research area.

Table 1: Summary statistics for variables used in the gravity model estimations

| Variable | Source | Obs. | Mean | S.D. | Min. | Max. |
|---|---|---|---|---|---|---|
| Number of publications | WoS | 25,344 | 19,392.1 | 34,473.9 | 32 | 383,875 |
| > Dual Use Category | WoS | 25,344 | 6,002.7 | 13,214.3 | 2 | 176,338 |
| > Technology | WoS | 25,344 | 5,042.9 | 11,892.3 | 3 | 167,986 |
| > Physical Sciences | WoS | 25,344 | 7,740.7 | 14,444.4 | 1 | 157,945 |
| > Life Science & Biology | WoS | 25,344 | 8,197.3 | 12,567.5 | 14 | 114,366 |
| > Social Science & Humanities | WoS | 25,344 | 1,308.4 | 2,630.9 | 0 | 22,409 |
| Number of co-publications | WoS | 25,344 | 328.58 | 800.04 | 0 | 12,403 |
| > Dual Use Category | WoS | 25,344 | 87.32 | 199.37 | 0 | 5973 |
| > Technology | WoS | 25,344 | 55.09 | 151.27 | 0 | 5670 |
| > Physical Sciences | WoS | 25,344 | 166.14 | 337.50 | 0 | 4579 |
| > Life Science & Biology | WoS | 25,344 | 130.44 | 339.36 | 0 | 5327 |
| > Social Science & Humanities | WoS | 25,344 | 12.49 | 43.63 | 0 | 899 |
| Co-publication rate (in %) | WoS | 25,344 | 1.27 | 1.57 | 0 | 18.37 |
| > Dual Use Category | WoS | 25,344 | 1.33 | 2.02 | 0 | 59.60 |
| > Technology | WoS | 25,344 | 0.85 | 1.16 | 0 | 15.86 |
| > Physical Sciences | WoS | 25,344 | 1.92 | 2.47 | 0 | 31.93 |
| > Life Science & Biology | WoS | 25,344 | 1.11 | 1.52 | 0 | 13.03 |
| > Social Science & Humanities | WoS | 25,344 | 0.58 | 1.28 | 0 | 68.29 |
| Per-capita GDP level, PPP | OECD | 25,344 | 28,562.6 | 15,706.8 | 2,036.8 | 97,864.2 |
| Geographical Distance in km | GW | 25,344 | 2,936.63 | 2,943.02 | 62 | 16,911 |

*Notes:* Per-capital GDP levels in Purchasing Power Parities (PPP) are measured in 2011 US Dollar prices; GW = Gleditsch and Ward (1999); an updated list of distances between country capitals is available at: http://ksgleditsch.com/data-5.html. For details on the definition of publication fields see main text and supplementary appendix.

***Stylized Facts.*** Before moving to the estimations, Panel A in Figure 1 provides stylized facts about the relative temporal development of international co-publication rates between the EU and Russia, on the one hand, and the EU and the remainder BRICS countries as control group, on the other hand. Data have been aggregated over EU and BRICS countries (averages). The visual inspection of the resulting two time series (bar graphs) serves as first eyeball regression for a potential change in co-publication rates before and after the economic sanctions were imposed in 2014. In Panel A of Figure 1, international co-publication rates have been normalised to 100% in 2013, the last sample year before the start of the sanctions. As the figure shows, co-publication rates in the years prior to 2013 were persistently higher for the EU-Russia relationship compared to the average international co-publication rate between the EU and the remainder BRICS. This changes from 2014 onwards, though, with average international co-publication rates between the EU and the remainder BRICS exceeding the corresponding rate for EU-Russia. This temporal coincidence between changes in relative co-publication rates and the imposition of sanctions can be taken as a first hint for their potential effects on knowledge flows.

Figure 1: EU-Russia co-publication rate relative to BRICS (aggregate) and a synthetic control unit

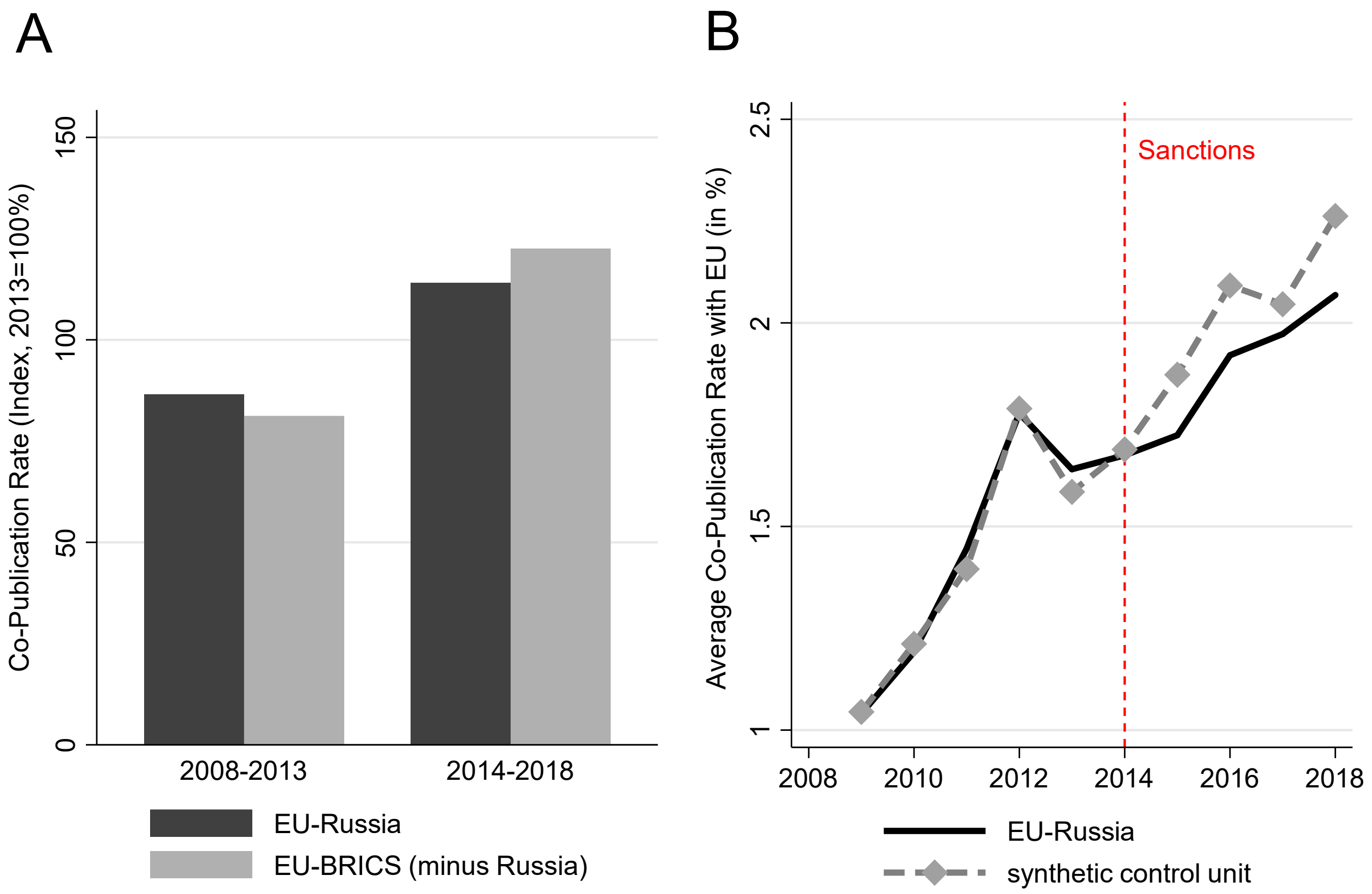

*Notes:* Comparison in Panel B based on Synthetic Control Method (SCM) applied to aggregate data for the average co-publication rate between sample countries and the EU-28. EU-Russia is the treated unit and the synthetic control unit has been selected which minimizes outcome differences in period 2008–2013 based on the following set of covariates: lagged co-publication rate, annual growth in co-publication rate, annual growth in overall publications, annual growth of per-capita GDP. Sample weights for the synthetic control unit have been estimated as follows: Slovakia 62.1%, UK 12.1%, India 21.1% and China 4.7% (all with [remainder] EU-28).

This impression is also confirmed in Panel B of Figure 1, which is based on a synthetic control method (SCM) approach as frequently used tool for comparative case studies of aggregate time-series data (Abadie et al., 2015; Abadie, 2020). The idea of SCM is to find a suitable synthetic control unit for EU-Russia that experiences as very similar development in average co-publication rates prior to the imposition of economic sanctions in 2014. This matched synthetic control unit is then used to track the relative development after the start of this treatment as a means for causal inference. As shown in Panel B of Figure 1, the temporal development in international co-publications rates for the estimated synthetic control unit is indeed very similar to the one for EU-Russia prior to 2014. After the start of sanctions, the figure reveals a widening gap in the development of international co-publication rates between the synthetic control unit and EU-Russia. Although estimation is based on few aggregate observations, the SCM results provide an additional indication for a potentially negative effect of sanctions on international knowledge flows. This link is investigated further based on individual (dyadic) data for country pairs and a rigorous panel econometric analysis next.

## 5. Empirical Results

***Aggregate Findings.*** Table 2 reports the estimated treatment effects for gravity model specifications as in Eq.(1) and extensions thereof. Specifications vary in terms of either including explicit control variables (such as geographical distance) or incorporating a multidimensional fixed effects structure (i.e. when we add country-pair fixed effects to the model this eliminates the geographical distance variable from the regressions). The motivation for displaying different specifications is to illustrate key model properties, on the one hand, and to provide robust treatment estimates of sanctions, on the other hand. Specifically, while the inclusion of geographical distance allows us to check if international co-publications follow an underlying gravity model logic, the use of multidimension fixed effects is better suited to circumvent endogeneity issues related to reversed causality and thus avoid estimation biases.

Accordingly, specification (I) places an emphasis on highlighting the working of the basic gravitational forces on international co-publications and thus places only very few restrictions on the data generating process in terms of including higher-order fixed effects to account for latent factors confounding with the treatment effects. The set of regressors include geographical distance, the product of per capita income and publication levels for country pairs together with binary dummies to account for systematic differences over time (i.e. after the start of the sanctions), cross-sectional differences (i.e. for EU-Russia country pairs) and combinations thereof (i.e., the key treatment variable *SANC* measuring the treatment effect of sanctions on the treatment group relative to the included control groups). As the results show, increasing geographical distance significantly reduces international co-publication rates between country pairs, thus, confirming prior research highlighting that knowledge flows are subject to a distance decay (Peri, 2005; Picci, 2010). While the results indicate that international co-publication rates have generally grown in 2014–2018 compared to previous years, the average co-publication rate between the EU and Russia is found to be generally lower compared to the set of control groups (intra-EU, EU-BRICS, Russia-BRICS). However, once we include country-fixed effects in specification (II) of Table 2 to capture structural differences in national innovation systems, the estimates indicate a higher average co-publication rate for EU-Russia country pairs *vis-à-vis* the control groups.

Most importantly, the gravity model estimates point to significantly negative effects of sanctions (*SANC*) on the treatment group. The average treatment effect on the treated is -0.00374 (*t*-value: -4.88, 95% CI: -0.00525, -0.00224) indicating that sanctions reduce international co-publications by approx. 0.37 %-points. If we relate this estimate to the average value of the international co-publication rate in the treatment group prior to the sanctions (1.08%), this implies a decline in international co-publication activity by approx. 34% caused by the sanctions. Treatment effects are – in tendency – larger if we account more precisely for unobserved confounding factors in specifications (II) to (V). For specifications (IV) and (V), which are considered to be robust to unobserved confounding factors as they include a complex set of fixed effects by controlling for latent, time-constant differences across country pairs (such as geographical distance, common language, historical and political ties etc.) and heterogeneous time-fixed effects at the country level, the reduction in the international co-publication activity is estimated to be 0.61 %-points (*t*-stat: -4.36, 95% CI: -0.00879, -0.00334) for the fixed effects (FE) linear panel model and 0.76 %-points (*t*-stat: -2.19, 95% CI: -0.0144, -0.000807) for the PPML estimator. This corresponds to a reduction in international co-publication activity by approx. 56% to 70%.

*Robustness Tests.* The estimation results presented in Table 3 further test the sensitivity of these results in two important dimensions. First, as shown in specification (II) in Table 3, the control group is reduced to EU-BRICS and Russia-BRICS country pairs by eliminating intra-EU country pairs. This sample reduction is motivated by the fact the intra-EU co-publication activities have intensified over time with the progress of European integration and may thus lead to an over-estimation of treatment effects of EU-Russia sanctions. Indeed, the estimated treatment effects for the subsample become smaller and amount to a 0.15 %-point reduction (*t*-stat: -1.96, CI: -0.00308, 0.0000135) for the sub-sample estimates; importantly, however, they stay significant at the 5% critical level. This effect translates into a decline in international co-publication activity by approx. 15% when comparing the EU-Russia to EU-BRICS (minus Russia) scientific knowledge collaboration.

Table 2: Gravity model estimates for treatment effects of EU-Russia sanctions on scientific co-publication rates

| Specification | (I) | (II) | (III) | (IV) | (V) |
|---|---|---|---|---|---|
| Estimator | POLS | POLS | FE | FE | PPML |
| Time dummy 2014–2018 | 0.00792*** | | | | |
| | (*t*-stat: 27.43) | | | | |
| | [CI: 0.00735, 0.00849] | | | | |
| Dummy Russia-EU Pairs | -0.00302*** | 0.00228*** | | | |
| | (*t*-stat: -9.84) | (*t*-stat: 4.32) | | | |
| | [CI: -0.00362, -0.00242] | [CI: 0.00125, 0.00332] | | | |
| Geographical Distance | -0.00629*** | -0.00915*** | | | |
| (*DIST*) | (*t*-stat: 54.21) | (*t*-stat: -43.83) | | | |
| | [CI: -0.00660, -0.00614] | [CI: -0.00956, -0.00874] | | | |
| Economic Sanctions | -0.00374*** | -0.00720*** | -0.00720*** | -0.00606*** | -0.00760** |
| (*SANC*) | (*t*-stat: -4.88) | (*t*-stat: -9.44) | (*t*-stat: -8.52) | (*t*-stat: -4.36) | (*t*-stat: -2.19) |
| | [CI: -0.00525, -0.00224] | [CI: -0.00869, -0.00570] | [CI: -0.00885, -0.00554] | [CI: -0.00879, -0.00334] | [CI: -0.0144, -0.000807] |
| Observations | 25,344 | 25,344 | 25,344 | 25,344 | 25,344 |
| Control Variables | Yes | Yes | Yes | Yes | Yes |
| Time FE | — | Yes | Yes | Yes | Yes |
| Country FE | — | Yes | — | — | — |
| Country-Pair FE | — | — | Yes | Yes | Yes |
| Country × Time FE | — | — | — | Yes | Yes |

*Notes:* * $p<0.1$; ** $p<0.05$; *** $p<0.01$. POLS = Pooled OLS, FE = Fixed Effects panel model, PPML = Poisson Pseudo Maximum Likelihood estimator; *t*-statistics in round brackets and 95% confidence intervals (CI) in square brackets with underlying standard errors are clustered at the country-pair level; sample period for estimation 1995–2018.

Table 3: Robustness tests for treatment effects of EU-Russia sanctions and pseudo-treatment effects

| Specification<br>Sample | (I)<br>All country pairs | (II)<br>Excluding intra-EU country pairs | (III)<br>All country pairs | (IV)<br>Excluding intra-EU country pairs | (V)<br>All country pairs | (VI)<br>Excluding intra-EU country pairs |
|---|---|---|---|---|---|---|
| *SANC* | -0.00606*** | -0.00153** | | | | |
| (Russia-EU × Time 2014–18) | (*t*-stat: -4.96) | (*t*-stat: -1.96) | | | | |
| | [CI: -0.00879, -0.00334] | [CI: -0.00308, 0.0000135] | | | | |
| *Pseudo-SANC I* | | | 0.00346*** | 0.00152** | | |
| (Russia-EU × Time 2000–07) | | | (*t*-stat: 4.43) | (*t*-stat: 2.59) | | |
| | | | [CI: 0.00193, 0.00500] | [CI: 0.000361, 0.00268] | | |
| *Pseudo-SANC II* | | | | | -0.001000 | -0.000886** |
| (Russia-EU × Time 2008–13) | | | | | (*t*-stat: -1.59) | (*t*-stat: -2.36) |
| | | | | | [CI: -0.00224, 0.000236] | [CI: -0.00163, -0.000143] |
| Observations | 25,344 | 3,820 | 25,344 | 3,820 | 25,344 | 3,820 |
| Controls Variables | Yes | Yes | Yes | Yes | Yes | Yes |
| Time FE | Yes | Yes | Yes | Yes | Yes | Yes |
| Country-Pair FE | Yes | Yes | Yes | Yes | Yes | Yes |
| Country × Time FE | Yes | — | Yes | — | Yes | — |

*Notes:* * $p<0.1$; ** $p<0.05$; *** $p<0.01$. Estimates are based on FE panel model with robust standard errors; *t*-statistics in round brackets and 95% confidence intervals in square brackets with underlying standard errors are clustered at the country-pair level; sample period for estimation 1995–2018.

Second, the remainder of the estimation results shown in Table 3 test for pseudo-treatment effects in different time intervals prior to the imposition of sanctions in 2014. If these pseudo-treatment effects would also be negative and significant, it would cast doubts on the causal nature of the revealed treatment effects after 2014 indicating that the effects would then simply follow a negative long-run trend. The estimation results for the time interval 2000–2007 indicate, however, that the development of the international co-publication rate for the treatment group EU-Russia is positive and larger than the corresponding trend development in the two control groups (all country pairs in specification (III), only EU-BRICS and Russia-BRICS country pairs in specification (IV)) relative to the baseline period 1995–1999. For 2008–2013 the estimation results show statistically insignificant (specification (V)) or very small negative results (specification (VI)). The latter may, though, point to early anticipation effects, which will be further explored in a dynamic treatment analysis.

Third, we have run a series of sub-sample estimates which subsequently leave out one EU country from the panel regressions. The logic if these "leave one out"-tests is to see if the negative sanctions impact on international scientific collaborations is not a universal treatment effect but driven by a drop in the co-publication rate between Russia and specific EU countries. As Table A2 in the supplementary appendix shows, however, the negative coefficient for our treatment indicator (SANC) remains negative and statistically significant for all "leave one out" regressions. This indicates that sanctions lead to a general reduction of scientific knowledge flows between the EU and Russia. Additionally, Table A3 shows "leave one out"-tests for individual BRICS countries to ensure that the negative treatment effect is not the result of a strong increase in the scientific collaboration between the EU and individual BRICS countries. As the table shows, leaving out individual BRICS also does not affect the direction and statistical significance of the estimated treatment effect.

***Scientific Fields.*** Table 4 reports average treatment effects for EU-Russia by scientific fields. The results for the full sample of all country pairs (specification (I)) and the subsample excluding intra-EU county pairs (specification (II)) point to significant negative effects of sanctions in most fields considered. Only for Life Science & Biomedicine the subsample estimates in specification (II) fail to establish significant effects. For the Dual Use Category, sanctions are found to reduce the international co-publication rate by between 0.3 and 0.75 %-points; evaluated against the pre-treatment average of 1.37% for the treatment group, this reduction translates into a relative decline in international co-publication activity of between 23% and 55% caused by the sanctions. The corresponding reductions for the WoS scientific fields are: Technology (between 12% and 57%), Physical Science (between 42% and 84%), Life Science & Biomedicine (84%, only statistically significant for the full sample of country pairs) and Social Sciences, Humanities & Arts (between 50% and 64%).

Table 4: Estimated treatment effects of EU-Russia sanctions by publication fields

| Specification<br>Treatment indicator *SANC* by publication field | (I)<br>All country pairs | (II)<br>Excluding intra-EU country pairs |
|---|---|---|
| Dual Use Category | -0.00755*** | -0.00311*** |
| | (*t*-stat: -5.86) | (*t*-stat: -4.05) |
| | [CI: -0.0101, -0.00502] | [CI: -0.00462, -0.00159] |
| Technology | -0.00556*** | -0.00112* |
| | (*t*-stat: -5.49) | (*t*-stat: -1.90) |
| | [CI: -0.00754, -0.00357] | [CI: -0.00229, 0.0000431] |
| Physical Sciences | -0.0151*** | -0.00754*** |
| | (*t*-stat: -4.81) | (*t*-stat: -6.03) |
| | [CI: -0.0212, -0.00892] | [CI: -0.0100, -0.00507] |
| Life Science & Biomedicine | -0.00574*** | 0.000478 |
| | (*t*-stat: -5.49) | (*t*-stat: 0.78) |
| | [CI: -0.00780, -0.00369] | [CI: -0.000733, 0.00169] |
| Social Sciences, Humanities & Arts | -0.00211*** | -0.00166*** |
| | (*t*-stat: -2.74) | (*t*-stat: -3.38) |
| | [CI: -0.00361, -0.000600] | [CI: -0.00264, -0.000690] |
| Observations | 25,344 | 3,820 |
| Controls (*GDP*, *Pub;* in logs) | Yes | Yes |
| Time FE | Yes | Yes |
| Country-Pair FE | Yes | Yes |
| Country × Time FE | Yes | — |

*Notes:* * $p<0.1$; ** $p<0.05$; *** $p<0.01$. Estimates are based on FE panel model; *t*-statistics in round brackets and 95% confidence intervals in square brackets with underlying standard errors are clustered at the country-pair level; sample period for estimation 1995–2018.

Figure 2: Estimated yearly treatment effects (overall and by scientific fields)

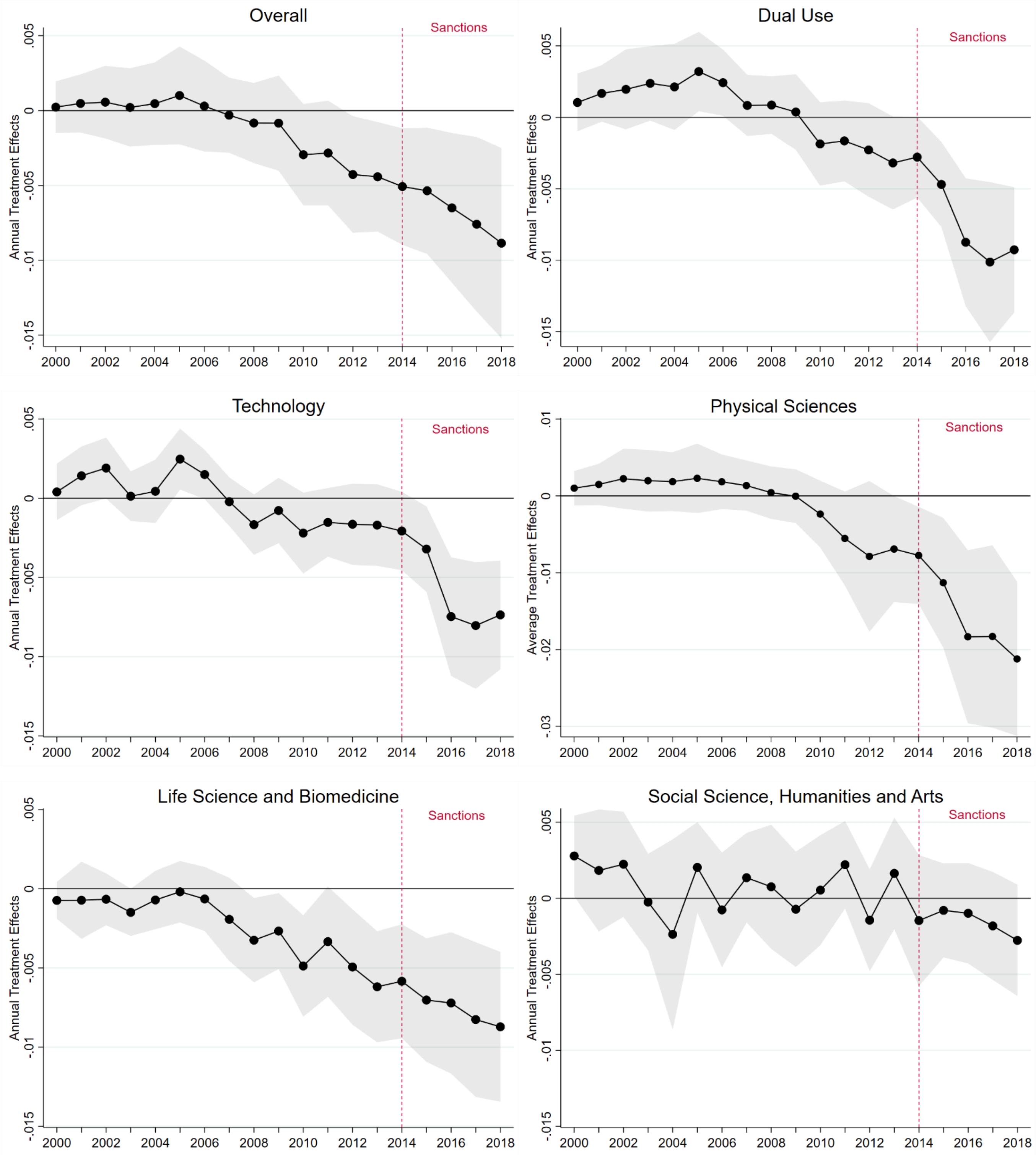

*Notes:* Grey areas indicate 99% confidence intervals for yearly treatment effects (black dot); underlying estimates are based on FE panel model including controls, time-, country-pair and country × time fixed effects; underlying standard errors have been clustered at the level of country-pairs; sample period for estimation 1995–2018 (all country pairs).

***Dynamic Specification.*** As another important sensitivity test for a negative sanctions effect, Figure 2 investigates the presence of dynamic treatment effects over time. Specifically, the figure plots annual coefficients for effect differences between the treatment and control group (all country pairs) relative to the baseline period 1995–1999 in a flexible DiD manner as shown in Eq.(2). In 3 out of 5 cases (by scientific fields), the timing of statistically significant negative treatment effects perfectly coincides with the imposition of sanctions in 2014. As an inspection of the individual panels of Figure 2 shows, this is the case for the Dual Use Category, Technology and Physical Sciences. While no systematic treatment effects can be detected prior to 2014 in these fields, effects turn persistently negative after the start of sanctions and grow over time. This latter observation indicates that the negative long-run consequences of sanctions on international scientific knowledge flows may even be larger than the reported average treatment effects in Table 4. For instance, in the case of the dual use category, treatment effects amount to 72% in 2017.

For Life Science & Biomedicine and the overall sample, the plotted annual coefficients in Figure 2 point to significant early anticipation effects, though. This indicates that in these cases the relative importance of EU-Russia scientific knowledge collaboration already started to decline throughout the late 2000s. In the case of Life Science & Biomedicine, this effect is mostly likely driven by deepening European integration in research and a resulting dynamically growing "Europeanisation" of the international co-publication activity within the EU (Mattsson et al., 2008) as it was already pointed out by the very heterogeneous point estimates for the two sample settings reported in Table 4. Importantly, however, the start of the sanctions shows to amplify negative effects – even if a tendency for early anticipation effects is visible in the flexible DiD estimation results. Finally, for Social Sciences, Humanities & Arts the annual coefficient plots point negative though statistically insignificant treatment effects when evaluated against the baseline period 1995–1999 and at a strict 99% confidence interval as applied in Figure 2.

## 6. Discussion and Conclusion

Since the start of the Ukraine crisis, the EU-Russia relations are in their most profound crisis since the end of the Cold War (Casier, 2020). Among the most controversially discussed policy tools in the current geo-political conflict are the EU's sanctions on the export of "dual use technologies" and the counter-sanctions of the Russian Federation, mainly hitting EU food exports. While earlier research has already highlighted the detrimental effects of economic sanctions and counter-sanctions on the movement of goods (i.e., trade), capital (e.g., FDI) and people, the question if sanctions have also affected the flows of knowledge between the EU and Russia has thus far remained an open one. We argue that a focus on international knowledge flows is well deserved given their role for the diffusion of new technologies and innovations as driver for economic growth and development at local and global scales (Coscia et al., 2020). Our results provide robust empirical evidence that the scientific co-publication activity between EU countries and the Russian Federation has significantly lost ground relative to the scientific collaboration activities for similar control groups, such as scientific co-publications between EU countries and the remainder BRICS economies since the imposition of sanctions in 2014.

We argue that this revealed temporal coincidence between the relative decline in EU-Russia co-publications and the start of reciprocal economic sanctions can likely be interpreted as causal. Although the measurement of the causal impact of state actions on socio-economic outcomes is surely one of the biggest methodological challenges in the field of international political science (Braumoeller et al., 2018), we argue that our chosen empirical identification approach carefully accounts for estimation challenges related to effect endogeneity and potential biases stemming from unobserved variables that confound the relationship between these political actions and outcomes. As one important aspect, we exploit the quasi-experimental nature of the rapid imposition of sanctions, which quickly entered into force after the Russian annexation of Crimea in March 2014 and thus provide an exogenous source of variation in our data.

In connection to this exogeneity argument, we have carefully accounted for confounding factors in the estimation of our gravity model specification by including multidimensional fixed effects capturing both time-fixed structural characteristics at the level of country pairs (such as geographical distance, common language, historical ties, etc.) and also time-varying macroregional factors other than the start of the economic sanctions. The selection of our control groups has been guided by the principles of 1) similarity in terms of pre-sanction scientific collaborations relative to the EU-Russia case and, importantly, 2) independence with regard to the imposition of sanctions on and from Russia (this is why, for instance, the USA has not been included in the sample). Finally, we have comprehensively tested for effect differences across scientific fields (including a tailored "Dual Use" Category) and for differently designed control groups (e.g. excluding intra-EU country pairs). Except for few scientific fields such as Social Sciences, Humanities and Arts (as shown in Figure 2), we find persistent negative effects of sanctions on knowledge flows.

What is less clear, however, is whether policy makers from involved conflict parties regard the revealed negative effects on knowledge flows as an additional dimension of sanction effectiveness or rather as an unintended side effect thereof. If the latter is the case, our results urge policy makers to consider these effects when assessing the overall benefits and costs of the imposed economic sanctions. In this context, our results may then also point at the importance of compensating policy actions (Casier, 2020) for securing international knowledge exchange even in times of political conflicts and sanctions. Such compensating policies would be especially important for sustaining scientific collaboration between the EU and Russia, since it takes time to establish mutual trust needed for building and maintaining successful inter-organisational and international scientific networks (Newell & Swan, 2000; Isabelle et al., 2011).

Cutting off international ties, even if sanctions only apply for a limited time window, may, thus, have long-lasting effects on such networks. Complementary research to our study has already shown that the Russian government is implementing a 'Turn to the East' -policy by strengthening their partnerships with countries such as China and South Korea to compensate for the loss of collaboration opportunities with Western countries (Shida, 2020). This development coincides with a significant increase in the domestic funding contribution for science in Russia (Schiermeier, 2018). As a result, it seems that the importance of the "EU connection" for Russian scholars has diminished.

Obviously, further research is needed to better understand whether such divergence trends following a geo-political crisis and sanctions only affect the relative geographical distribution of international knowledge exchange or whether they also result in an allocational loss (collateral damage) in terms of knowledge creation for involved conflict parties. Regarding the specific EU-Russia relationship, further research on the topic is called for since there seems to be no end in sight for the sanctions. Quite the contrarily, the latest incidents concerning the opposition politician Alexei Navalny and his supporters have prompted EU and USA to impose additional sanctions against Russia. Still, while at the moment the lifting of the sanctions seems unlikely to happen in the near future, it would be also be vital to gain a better understanding of what will happen if the sanctions are terminated (see, e.g., Attia et al., 2020; Crozet et al., 2021): Will scientific collaboration return to the "pre-sanctions level" or will there be a permanent drop in the level of EU-Russia collaboration as new permanent networks have been built up? Such potential path dependencies from economic sanctions should be investigated once longer time series on scientific collaborations and knowledge flows are available.

## References

Abadie, A., 2020. Using synthetic controls: Feasibility, data requirements, and methodological aspects. *Journal of Economic Literature* https://www.aeaweb.org/articles?id=10.1257/jel.20191450

Abadie, A., Diamond, A. & Hainmueller, J., 2015. Comparative politics and the synthetic control method. *American Journal of Political Science* 59, 495–510.

Acosta, M., Coronado, D., Ferrándiz, E., Marín, M. R., & Moreno, P. J., 2020. Civil–military patents and technological knowledge flows into the leading defense firms. *Armed Forces & Society*46, 454–474.

Ahn, D. P. & Ludema, R. D., 2020. The sword and the shield: The economics of targeted sanctions. *European Economic Review* 130, 103587.

Aldieri, L., Kotsemir, M. & Vinci, C. P., 2018. The impact of research collaboration on academic performance: An empirical analysis for some European countries. *Socio-Economic Planning Sciences* 62, 13–30.

Aman, V., 2018. A new bibliometric approach to measure knowledge transfer of internationally mobile scientists. *Scientometrics* 117, 227–247.

Appelt, S., van Beuzekom, B., Galindo-Rueda, F., & de Pinho, R., 2015. Which factors influence the international mobility of research scientists? In A. Geuna (Ed.): *Global mobility of research scientists* (pp. 177–213). Academic Press: Cambridge.

Atta-Owusu, K., 2019. Oasis in the desert? Bridging academics' collaboration activities as a conduit for global knowledge flows to peripheral regions. *Regional Studies, Regional Science* 6, 265–280.

Attia, H., Grauvogel, J. & von Soest, C., 2020. The termination of international sanctions: Explaining target compliance and sender capitulation. *European Economic Review,* 129, 103565.

Aykac, G., 2021. The value of an overseas research trip. *Scientometrics* 126, 7097–7122.

Baier, S. L. & Bergstrand, J. H., 2007. Do free trade agreements actually increase members' international trade? *Journal of International Economics* 71, 72–95.

Balazsi, L., Matyas, L. & Wansbeek, T., 2018. The estimation of multidimensional fixed effects panel data models. *Econometric Reviews* 37, 212–227.

Bělín, M. & Hanousek, J., 2020. Which sanctions matter? Analysis of the EU/Russian sanctions of 2014. *Journal of Comparative Economics* https://doi.org/10.1016/j.jce.2020.07.001

Besedeš, T., Goldbach, S. & Nitsch, V., 2021. Cheap talk? Financial sanctions and non-financial firms. *European Economic Review* 134, 103688.

Braumoeller, B., Marra, G., Radice, R. & Bradshaw, A., 2018. Flexible causal inference for political science. *Political Analysis* 26, 54–71.

Caloghirou, Y., Vonortas, N. S., & Ioannides, S., 2006. Knowledge flows in European industry. Edward Elgar, London.

Caldara, D. & Iacoviello, M., 2018. Measuring Geopolitical Risk. International Finance Discussion Papers No.1222. Available at (accessed 28-11-2021): https://doi.org/10.17016/IFDP.2018.1222.

Callaway, B. & Sant'Anna, P. 2020. Difference-in-Differences with multiple time periods. *Journal of Econometrics*, in press. https://doi.org/10.1016/j.jeconom.2020.12.001

Casier, T., 2020. Not on speaking terms, but business as usual: The ambiguous coexistence of conflict and cooperation in EU-Russia relations. *East European Politics* 36, 529–543.

Chen, K., Zhang, Y. & Fu, X., 2019. International research collaboration: An emerging domain of innovation studies? *Research Policy* 48, 149–168.

Ciani, E. & Fisher P., 2019. Dif-in-Dif estimators of multiplicative treatment effects. *Journal of Econometric Methods* 8, 1–10

Coscia, M. & Wang, L., 2016. Evolution and convergence of the patterns of international scientific collaboration. *Proceedings of the National Academy of Sciences* 113, 2057–2061.

Coscia, M., Neffke, F. M. H. & Hausmann, R., 2020. Knowledge diffusion in the network of international business travel. *Nature Human Behaviour* 4, 1011–1020.

Crozet, M. & Hinz, J., 2020. Friendly fire: The trade impact of the Russia sanctions and counter-sanctions. *Economic Policy* 35, 97–146

Crozet, M., Hinz, J., Stammann, A. & Wanner, J., 2021. Worth the pain? Firms' exporting behaviour to countries under sanctions. *European Economic Review* 134, 103683.

Dai, M., Yotov, Y. V. & Zylkin, T., 2014. On the trade-diversion effects of free trade agreements. *Economics Letters* 122, 321–325.

Di Cagno, D., Fabrizi, A., & Meliciani, V., 2014. The impact of participation in European joint research projects on knowledge creation and economic growth. *Journal of Technology Transfer* 39, 836–858.

Di Cagno, D., Fabrizi, A., Meliciani, V., & Wanzenböck, I., 2016. The impact of relational spillovers from joint research projects on knowledge creation across European regions. *Technological Forecasting and Social Change* 108, 83–94

Doornich, J. B. & Raspotnik, A., 2020. Economic sanctions disruption on international trade patterns and global trade dynamics: Analyzing the effects of the European Union's sanctions on Russia. *Journal of East-West Business* 26, 344–364

Dreger, C., Kholodilin, K. A., Ulbricht, D. & Fidrmuc, J., 2016. Between the hammer and the anvil: The impact of economic sanctions and oil prices on Russia's ruble. *Journal of Comparative Economics* 44, 295–308.

Education Statistics Finland, 2021. Kansainvälisyys. https://vipunen.fi/fi-fi/

European Commission, 2018. Roadmap for EU-Russia S&T Cooperation. https://ec.europa.eu/info/sites/default/files/research_and_innovation/strategy_on_research_and_innovation/documents/ec_rtd_russia-roadmap_2018.pdf

European Council, 2021. Timeline – EU Restrictive Measures in Response to the Crisis in Ukraine. https://www.consilium.europa.eu/en/policies/sanctions/ukraine-crisis/history-ukraine-crisis/

Fally, T., 2015. Structural gravity and fixed effects. *Journal of International Economics* 97, 76–85.

Felbermayr, G., Kirilakha, A., Syropoulos, C., Yalcin, E. & Yotov, Y. V., 2020. The global sanctions data base. *European Economic Review* 129, 103561.

Felbermayr, G., Morgan, T. C., Syropoulos, C. & Yotov, Y. V., 2021. Understanding Economic Sanctions: Interdisciplinary Perspectives on Theory and Evidence. *European Economic Review* 135, 103720.

Fernández-Zubieta, A., & Guy, K., 2010. *Developing the European Research Area: Improving knowledge flows via researcher mobility.* JRC Scientific and Technical Report, JRC-IPTS.

Gibson, J., & McKenzie, D., 2014. Scientific mobility and knowledge networks in high emigration countries: Evidence from the Pacific. *Research Policy* 43, 1486–1495.

Giumelli, F., 2017. The redistributive impact of restrictive measures on EU members: Winners and losers from imposing sanctions on Russia. *Journal of Common Market Studies* 55, 1062–1080.

Gleditsch, K. S. & Ward, M. D., 1999. Interstate system membership: A revised list of the independent states since 1816. *International Interactions* 25, 393–413.

Gui, Q., Liu, C. & Du, D., 2018. Does network position foster knowledge production? Evidence from international scientific collaboration network. *Growth and Change* 49, 594–611.

Hovi, J., Huseby, R. & Sprinz, D. F., 2005. When do (imposed) economic sanctions work? *World Politics* 57, 479–499.

Hufbauer, G. C. & Jung, E., 2020. What's new in economic sanctions? *European Economic Review* 130, 103572.

Jöns, H., 2009. 'Brain circulation'and transnational knowledge networks: studying long-term effects of academic mobility to Germany, 1954–2000. *Global Networks* 9, 315–338.

Isabelle, D. A. & Heslop, L. A., 2011. Managing for success in international scientific collaborations: Views from Canadian government senior science managers. *Science and Public Policy* 38, 349–364.

Ivanov, S., Sypchenko, L. & Webster, C., 2017. International sanctions and Russia's hotel industry: The impact on business and coping mechanisms of hoteliers. *Tourism Planning & Development* 14, 430–441.

Keller, W. & Yeaple, S., 2013. The gravity of knowledge. *American Economic Review* 103, 1414–1444.

Kotsemir, M., Kuznetsova, T., Nasybulina, E. & Pikalova, A., 2015. Identifying directions for Russia's science and technology cooperation. *Foresight and STI Governance* 9, 54–72.

Krige, J., 2014. National security and academia: Regulating the international circulation of knowledge. *Bulletin of the Atomic Scientists* 70, 42–52.

Liuhto, K., Sutyrin, S. & Blanchard, J-M., 2017. *The Russian Economy and Foreign Direct Investment.* Routledge, Abingdon.

Lovakov, A. & Agadullina, E., 2019. Bibliometric analysis of publications from post-Soviet countries in psychological journals in 1992–2017. *Scientometrics* 119, 1157–1171.

Makkonen, T. & Mitze, T., 2016. Scientific collaboration between 'old' and 'new' member states: Did joining the European Union make a difference? *Scientometrics* 106, 1193–1215.

Makkonen, T., Williams, A. M., Weidenfeld, A. & Kaisto, V., 2018. Cross-border knowledge transfer and innovation in the European neighbourhood: Tourism cooperation at the Finnish–Russian border. *Tourism Management* 68, 140–151.

Mattsson, P., Laget, P., Nilsson, A. & Sundberg, C. J., 2008. Intra-EU vs. extra-EU scientific co-publication patterns in EU. *Scientometrics* 75, 555–574.

McCullagh, P. & Nelder, J., 1989. *Generalized Linear Models.* Chapman and Hall, London.

Melin, G. & Persson, O., 1996. Studying research collaboration using co-authorships. *Scientometrics* 36, 363–377

Nelson, A., 2009. Measuring knowledge spillovers: What patents, licenses and publications reveal about innovation diffusion. *Research Policy* 38, 994–1005.

Newell, S. & Swan, J., 2000. Trust and inter-organizational networking. *Human Relations* 53, 1287–1328.

Newman, M. E., 2004. Coauthorship networks and patterns of scientific collaboration. *Proceedings of the National Academy of Sciences* 101, 5200–5205.

Peri, G., 2005. Determinants of knowledge flows and their effect on innovation. *Review of Economics and Statistics* 87, 308–322.

Petersen, A. M., 2018. Multiscale impact of researcher mobility. *Journal of the Royal Society Interface,* 15, 20180580.

Picci, L., 2010. The internationalization of inventive activity: A gravity model using patent data. *Research Policy* 39, 1070–1081.

Prokkola, E-K., 2019. Border-regional resilience in EU internal and external border areas in Finland. *European Planning Studies* 27, 1587–1606.

Rose, A. K., 2004. Do we really know that the WTO increases trade? *American Economic Review* 94, 98–114.

Santos Silva, J. M. C. & Tenreyro, S., 2006. The log of gravity. *Review of Economics and Statistics* 88, 641–658.

Scalia, T. et al., 2017. *Final Technical Report: Study on the Dual-Use Potential of Key Enabling Technologies.* Brussels: European Union.

Schiermeier, Q., 2018. Russian science chases escape from mediocrity. *Nature* 555, 297–298.

Schiller, D., & Diez, J. R., 2012. The impact of academic mobility on the creation of localized intangible assets. *Regional Studies* 46, 1319–1332.

Shida, Y., 2020. Russian business under economic sanctions: Is there evidence of regional heterogeneity? *Post-Communist Economies* 32, 447–467.

Trippl, M., 2013. Scientific mobility and knowledge transfer at the interregional and intraregional level. *Regional Studies* 47, 1653–1667.

Ubfal, D., & Maffioli, A., 2011. The impact of funding on research collaboration: Evidence from a developing country. *Research Policy,* 40(9), 1269-1279.

Volovik, N., 2020. Russia-EU trade development under the sanctions. *Baltic Rim Economies* 2020, 2738.

Wagner, C., Whetsell, T. & Mukherjee, S., 2019. International research collaboration: Novelty, conventionality, and atypicality in knowledge recombination. *Research Policy* 48, 1260–1270.

Wooldridge, J., 2010. *Econometric Analysis of Cross Section and Panel Data.* MIT Press, Cambridge.

Yegorov, I., 2009. Post-Soviet science: Difficulties in the transformation of the R&D systems in Russia and Ukraine. *Research Policy* 38, 600–609.

Zhang, Z., Rollins, J. & Lipitakis, E., 2018. China's emerging centrality in the contemporary international scientific collaboration network. *Scientometrics* 116, 1075–1091.

**Supplementary Appendix**

Table A1: Matching between dual-use technologies and Web of Science subject categories

| **Dual-use technology** | **Web of Science subject categories** |
|---|---|
| Energy for mobility<br>(e.g., hydrogen storage systems for fuel cells) | 1. Electrochemistry<br>2. Energy & Fuels<br>3. Engineering, Chemical<br>4. Engineering, Mechanical<br>5. Engineering, Petroleum<br>6. Transportation Science & Technology |
| Fundamental non-dependence materials and components<br>(e.g., advanced, smart materials for satellite applications) | 7. Materials Science, Ceramics<br>8. Materials Science, Coating & Films<br>9. Materials Science, Composites<br>10. Metallurgy & Metallurgical Engineering<br>11. Nanoscience & Nanotechnology |
| Health and sanitary protection<br>(e.g., screening devices for detection of chemical, biological, radiological, nuclear and explosive material traces) | 12. Biotechnology & Applied Microbiology<br>13. Cell Biology<br>14. Chemistry, Inorganic & Nuclear<br>15. Engineering, Biomedical<br>16. Medical Laboratory Technology<br>17. Microscopy<br>18. Neuroimaging<br>19. Nuclear Science & Technology<br>20. Pharmacology & Pharmacy<br>21. Physics, Nuclear<br>22. Radiology, Nuclear Medicine & Medical imaging<br>23. Spectroscopy<br>24. Toxicology<br>25. Virology |
| Communication, navigation and surveillance systems<br>(e.g., unmanned vehicles for wide-area surveillance in air, land and water) | 26. Engineering, Aerospace<br>27. Engineering, Electrical & Electronic<br>28. Engineering, Marine<br>29. Engineering, Ocean<br>30. Imaging Science & Photographic Technology<br>31. Optics<br>32. Quantum Science & Technology<br>33. Remote Sensing<br>34. Telecommunications |

Table A1 (cont.): Matching between dual-use technologies and Web of Science subject categories

| | |
|---|---|
| Communication, navigation and surveillance systems (e.g., unmanned vehicles for wide-area surveillance in air, land and water) | 35. Engineering, Aerospace<br>36. Engineering, Electrical & Electronic<br>37. Engineering, Marine<br>38. Engineering, Ocean<br>39. Imaging Science & Photographic Technology<br>40. Optics<br>41. Quantum Science & Technology<br>42. Remote Sensing<br>43. Telecommunications |
| Human assistance and robotics (e.g., robotic systems such as exoskeletons) | 44. Anatomy & Morphology<br>45. Cell & Tissue Engineering<br>46. Materials Science, Biomaterials<br>47. Robotics |
| Security/cybersecurity systems (e.g., advanced biometric identification techniques) | 48. Computer Science, Artificial Intelligence<br>49. Computer Science, Cybernetics |
| Production and supply chain solutions (e.g., tracking and tracing devices) | 50. Automation & Control Systems<br>51. Engineering, Industrial<br>52. Engineering, Manufacturing<br>53. Engineering, Civil<br>54. Instruments & Instrumentation<br>55. Mechanics |

*Source:* Descriptions of dual use technologies is based on Scalia et al.'s (2017) work.

Figure A1: Teaching and research staff mobility between Finland and Russia in 2013–2019.

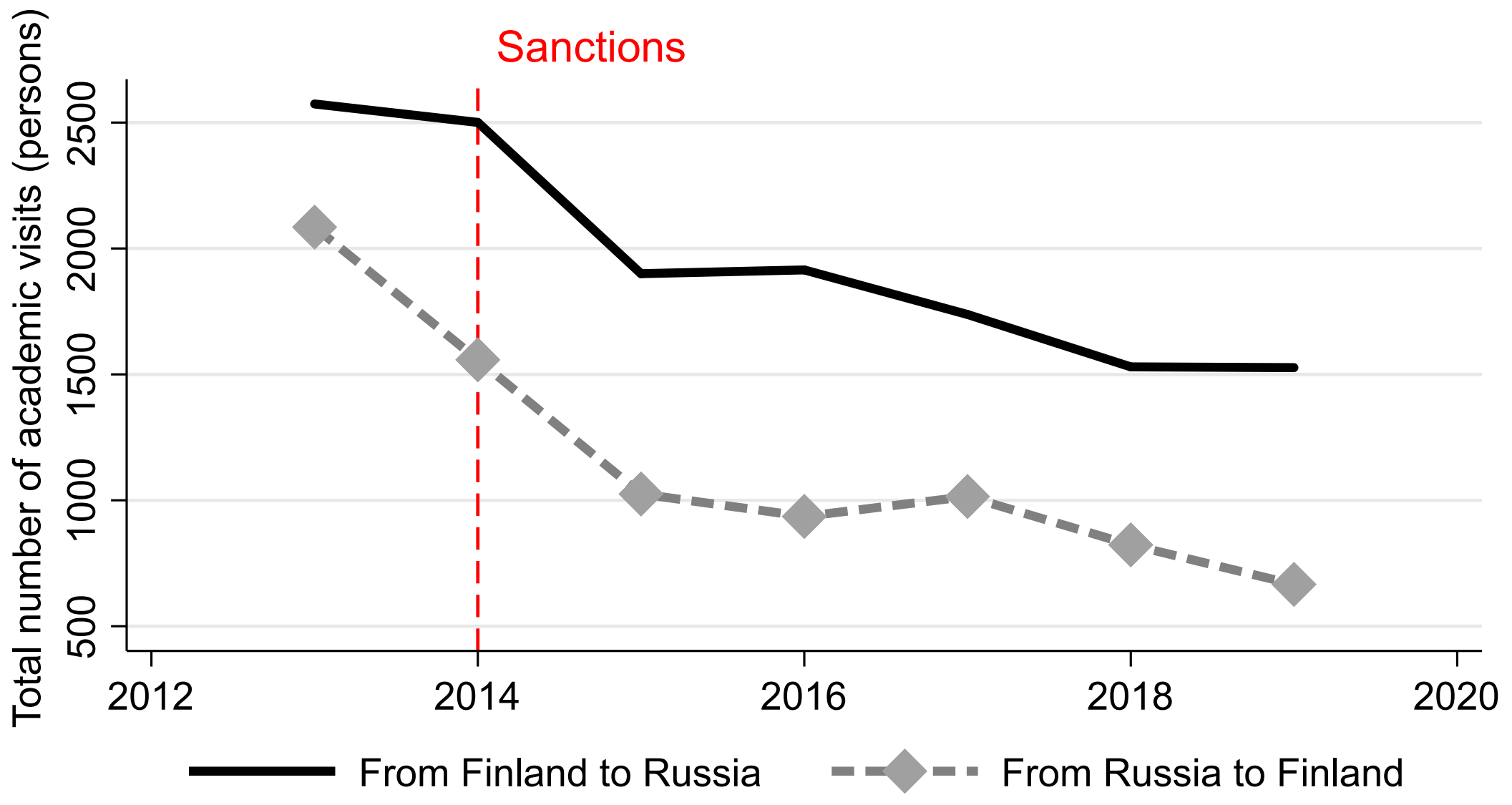

*Source:* Education Statistics Finland (2021).

Table A2: Leave-one-out robustness checks for impact of sanctions on knowledge flows (EU)

| Leave out country ↓ | Estimated coefficient for: Economic Sanctions (*SANC*) | | |
|---|---|---|---|
| Austria | -0.00600*** | (*t*-stat: 4.21) | [CI: -0.00880, -0.00320] |
| Belgium | -0.00597*** | (*t*-stat: 4.19) | [CI: -0.00877, -0.00318] |
| Bulgaria | -0.00568*** | (*t*-stat: 4.04) | [CI: -0.00844, -0.00292] |
| Croatia | -0.00566*** | (*t*-stat: 3.96) | [CI: -0.00846, -0.00285] |
| Cyprus | -0.00568*** | (*t*-stat: 4.03) | [CI: -0.00844, -0.00291] |
| Czech Republic | -0.00615*** | (*t*-stat: 4.34) | [CI: -0.00893, -0.00337] |
| Denmark | -0.00609*** | (*t*-stat: 4.28) | [CI: -0.00888, -0.00330] |
| Estonia | -0.00556*** | (*t*-stat: 3.94) | [CI: -0.00833, -0.00279] |
| Finland | -0.00622*** | (*t*-stat: 4.39) | [CI: -0.00900, -0.00344] |
| France | -0.00586*** | (*t*-stat: 4.21) | [CI: -0.00859, -0.00313] |
| Germany | -0.00598*** | (*t*-stat: 4.19) | [CI: -0.00878, -0.00318] |
| Greece | -0.00583*** | (*t*-stat: 4.11) | [CI: -0.00862, -0.00305] |
| Hungary | -0.00591*** | (*t*-stat: 4.17) | [CI: -0.00869, -0.00312] |
| Ireland | -0.00605*** | (*t*-stat: 4.22) | [CI: -0.00886, -0.00324] |
| Italy | -0.00599*** | (*t*-stat: 4.18) | [CI: -0.00881, -0.00318] |
| Latvia | -0.00573*** | (*t*-stat: 4.05) | [CI: -0.00851, -0.00295] |
| Lithuania | -0.00567*** | (*t*-stat: 3.99) | [CI: -0.00845, -0.00288] |
| Luxembourg | -0.00639*** | (*t*-stat: 4.44) | [CI: -0.00922, -0.00357] |
| Malta | -0.00638*** | (*t*-stat: 4.42) | [CI: -0.00921, -0.00354] |
| Netherlands | -0.00600*** | (*t*-stat: 4.31) | [CI: -0.00874, -0.00327] |
| Poland | -0.00626*** | (*t*-stat: 4.36) | [CI: -0.00908, -0.00344] |
| Portugal | -0.00629*** | (*t*-stat: 4.48) | [CI: -0.00905, -0.00354] |
| Romania | -0.00632*** | (*t*-stat: 4.50) | [CI: -0.00907, -0.00357] |
| Slovakia | -0.00619*** | (*t*-stat: 4.35) | [CI: -0.00898, -0.00340] |
| Slovenia | -0.00604*** | (*t*-stat: 4.23) | [CI: -0.00884, -0.00324] |
| Spain | -0.00621*** | (*t*-stat: 4.37) | [CI: -0.00899, -0.00342] |
| Sweden | -0.00604*** | (*t*-stat: 4.24) | [CI: -0.00883, -0.00324] |
| UK | -0.00694*** | (*t*-stat: 6.32) | [CI: -0.00909, -0.00479] |

*Notes:* * $p<0.1$; ** $p<0.05$; *** $p<0.01$. Estimated coefficients for Fixed Effects (FE) panel model; *t*-statistics in round brackets and 95% confidence intervals (CI) in square brackets with underlying standard errors are clustered at the country-pair level; sample period for estimation 1995–2018.

Table A3: Leave-one-out robustness checks for impact of sanctions on knowledge flows (BRICS)

| Leave out country ↓ | Estimated coefficient for: Economic Sanctions (*SANC*) | | |
|---|---|---|---|
| Brazil | -0.00650*** | (*t*-stat: -3.80) | [CI: -0.00986, -0.00315] |
| India | -0.00552*** | (*t*-stat: -3.87) | [CI: -0.00832, -0.00272] |
| China | -0.00618*** | (*t*-stat: -3.47) | [CI: -0.00967, -0.00269] |
| South Africa | -0.00691*** | (*t*-stat: -5.25) | [CI: -0.00949, -0.00433] |

*Notes:* * $p<0.1$; ** $p<0.05$; *** $p<0.01$. Estimated coefficients for Fixed Effects (FE) panel model; *t*-statistics in round brackets and 95% confidence intervals (CI) in square brackets with underlying standard errors are clustered at the country-pair level; sample period for estimation 1995–2018.